\documentclass[sigconf]{acmart}

\AtBeginDocument{%
  \providecommand\BibTeX{{%
    \normalfont B\kern-0.5em{\scshape i\kern-0.25em b}\kern-0.8em\TeX}}}

\renewcommand\footnotetextcopyrightpermission[1]{} 
\usepackage{lipsum}  

\usepackage[utf8]{inputenc}
\usepackage{graphicx}
\graphicspath{ {plots/} }
\usepackage{subfig}
\usepackage{amssymb}
\usepackage{amsmath}
\usepackage{amsthm}
\usepackage{caption}
\usepackage{bbm}
\usepackage{array,longtable}
\usepackage{times}
\usepackage{crossreference}
\usepackage{tikz}
\usepackage{booktabs}
\usepackage{enumitem}
\usepackage{eurosym}
\usepackage{array}
\usepackage{multirow}
\usepackage{hyphenat}

\newcommand{\blue}[1]{{#1}}

\theoremstyle{definition}
\theoremstyle{definition}
\newtheorem{experiment}{Definition}

\newcommand{\descr}[1]{\vspace{0.2cm} \noindent \textbf{#1}}

\newcommand{\sample}[1]{\mathrel{{\leftarrow}\vcenter{\hbox{\scriptsize\rmfamily\upshape\ensuremath{{#1}}}}}}

\newcommand{\adversary}{\mathcal{A}}

\renewcommand{\textrightarrow}{$\rightarrow$}

\begin{document}

\title{On Inferring Training Data Attributes in Machine Learning Models}

\author{Benjamin Zi Hao Zhao}
\email{benjamin.zhao@unsw.edu.au}
\affiliation{%
  \institution{University of New South Wales and Data61 CSIRO}
}
\author{Hassan Jameel Asghar}
\email{ hassan.asghar@mq.edu.au }
\affiliation{%
  \institution{Macquarie University and Data61 CSIRO}
}
\author{Raghav Bhaskar}
\email{Raghav.Bhaskar@data61.csiro.au}
\affiliation{%
  \institution{Data61 CSIRO}
}
\author{Mohamed Ali Kaafar}
\email{dali.kaafar@mq.edu.au}
\affiliation{%
  \institution{Macquarie University and Data61 CSIRO}
}


\begin{abstract}
A number of recent works have demonstrated that API access to machine learning models leaks information about the dataset records used to train the models.
Further, the work of \cite{somesh-overfit} shows that such \emph{membership inference} attacks (MIAs) may be sufficient to construct a stronger breed of \emph{attribute inference} attacks (AIAs), which given a partial view of a record can guess the missing attributes. In this work, we show (to the contrary) that MIA may not be sufficient to build a successful AIA. This is because the latter requires the ability to distinguish between similar records (differing only in a few attributes), and, as we demonstrate, the current breed of MIA are unsuccessful in distinguishing member records from \emph{similar} non-member records. We thus propose a relaxed notion of AIA, whose goal is to only approximately guess the missing attributes and argue that such an attack is more likely to be successful, if MIA is to be used as a subroutine for inferring training record attributes.

\end{abstract}




\keywords{Membership inference; Attribute inference; Neural networks.}


\maketitle

\section{Introduction}
The introduction of low-cost machine learning APIs from Google, Microsoft, Amazon, IBM, etc., has enabled many companies to monetize advanced machine learning models trained on private datasets by exposing them as a service. This has also caught the interest of the privacy community who has shown that these models may leak information about the records of the training dataset via membership inference attacks (MIA). In MIA, the adversary (a user of the service) may have simple black-box access to these models, submitting inputs (records) and obtaining outputs (class labels and probability/confidence for each label), can infer whether its target input was part of the training dataset or not. This can be a serious privacy breach when the underlying dataset is sensitive, e.g., medical data. 

The main reason for the success of an MIA is attributed to the observation that machine learning algorithms tend to return higher confidence scores for examples that they have seen (i.e., records in the training dataset) versus those that they encounter for the first time~\cite{shokri-mia, ml-leaks}. Overfitted models are shown to be more prone to MIA~\cite{shokri-mia, somesh-overfit}, although the attack has also been observed on well-generalized models~\cite{long-well-generalized}. It has also been observed that the attack's success rate varies on different \emph{sub-groups} of the training dataset~\cite{disparate-sub-group}. Some researchers have also investigated a related, and perhaps a more likely attack in practice, where the adversary only knows a partial record of a target in the training dataset and seeks to complete its knowledge of the missing attributes by accessing the machine learning model. This is known as the \emph{Attribute Inference Attack} (AIA)~\cite{somesh-overfit}. Samuel et al.~\cite{somesh-overfit} provide a formal definition of an AIA, and argue that an attribute inference adversary can infer the missing attribute value by using a membership inference adversary as a subroutine. Slightly more precisely, for a missing attribute with $t$ possible values, the AI adversary constructs $t$ different input (feature) vectors, gives them as input to the MI adversary, and outputs the attribute value which corresponds to the output from the MI adversary with the highest confidence~\cite[\S 5, p. 277]{somesh-overfit}. 

While this appears to be a straightforward result, in this paper we show that it is not true in practice. More specifically, we hypothesize and experimentally validate that an MI adversary is unable to effectively distinguish between a member of the training dataset and any non-members that are \emph{close} to the member according to some distance metric. This has implications for attribute inference, since the AI adversary would like to confidently infer the missing attribute value of its target record (who is assumed to be a member of the training dataset) by distinguishing it from other values (which result in vectors which are non-members). Thus, while machine learning algorithms output higher confidence for examples that are part of the training dataset, they also similarly return higher confidence scores for examples that are \emph{similar} to the training examples. Summary of our main contributions follow.

\begin{itemize}[topsep=1pt]
\item We propose the notion of \emph{strong} membership inference in which the adversary is asked to distinguish between two input vectors (records) which are a certain distance apart from each other (according to a given distance metric), where exactly one of the two is a member. We hypothesize and experimentally validate that as long as the vectors are \emph{close}, the adversary's advantage is not significantly better than random guess, and the advantage improves as the distance between the vectors increases. Since this implies that an attribute inference adversary is unlikely to succeed (if the number of missing attributes are small), we propose an alternative definition of \emph{approximate} attribute inference, where the goal of the adversary is to output a nearby (complete) vector to the target vector. 

\item We train neural network target models for MIA on the datasets of Location-30, and Purchases-2,10,20,50,100. The aim is to see if the confidence values between members and non-members can be distinguished. We replicate MIA accuracies reported in literature, and show that where MIA performs well, typically, the non-members (in the test datasets) have high Hamming distance from the nearest members. We then interpolate this by generating synthetic generated non-members (controlling the Hamming distance from members), and show that the accuracy of the distinguisher is similar to a random guess for vectors close to any member of the training set, while the accuracy grows as we move away from (any) members.




\end{itemize}

We remark that our definitions and experiments do not take the class labels into account (instead relying on the largest confidence value only). As a result, our attribute inference attack is not an instance of \emph{model inversion}~\cite{pharma-inversion}, which in essence relies on the correlation between attributes in the training dataset and the (learned) class label. 
A more detailed discussion on the merits of model inversion and similar attacks can be found in~\cite{shokri-mia}. 


\section{Definitions, Membership and Attribute Inference}
\label{sec:definition}
\descr{Notations.}
A machine learning model $f$ takes as input feature vectors $\mathbf{x} \in \mathbb{R}^m$ (of $m$ elements/features) and outputs a \emph{label vector} $\mathbf{y} \in \mathbb{R}^*$, whose $i$th element denotes the confidence (or probability) score for the $i$th class label. 
We denote by $\mathcal{P}$ the distribution induced by feature vectors $\mathbf{x}$ on $\mathbb{R}^m$. This captures the a priori belief about the feature space. A training dataset $D$ is a multiset of feature vectors from $\mathbb{R}^m$. 
In general, the rows of $D$ may be sampled according to some distribution other than $\mathcal{P}$. The notation $a \sample{\mathcal{D}} A$ indicates sampling an element $a$ from some set $A$ with some distribution $\mathcal{D}$. The symbol `$\$$' denotes uniform distribution. We denote by $f_D$, the model $f$ trained on the training dataset $D$. 

\descr{Distance, Neighbors and Portions.} Let $d$ be a metric on $\mathbb{R}^m$. For $\mathbf{x} \in \mathbb{R}^m$, the set $\mathsf{ngb}_r(\mathbf{x})$ denotes the set of all $r$-neighbors of $\mathbf{x}$, i.e., vectors $\mathbf{x}'$ such that $d(\mathbf{x}, \mathbf{x}') = r$, for a real number $r > 0$. The distance of a vector $\mathbf{x}' \in \mathbb{R}^m$ from a dataset $D$ is defined as $\min_{\mathbf{x} \in D} d(\mathbf{x}, \mathbf{x}')$.
For a vector $\mathbf{x}$, a \emph{portion} of $\mathbf{x}$, denoted $\mathbf{x}^*$, is a vector which has at least one feature of $\mathbf{x}$ replaced with $*$. 
The set of features which are \emph{starred} in $\mathbf{x}^*$ is known as the \emph{unknown part} of $\mathbf{x}^*$. We denote this as $\phi(\mathbf{x}^*)$. Each feature in $\phi(\mathbf{x}^*)$ is called an unknown feature. Without loss of generality we will assume that the unknown part of $\mathbf{x}^*$ consists of the last $|\phi(\mathbf{x}^*)|$ features.
Throughout the rest of this paper, with a focus on binary datasets, we shall use the Hamming distance as the instance of the distance metric $d$.

\descr{Membership and Attribute Inference.} We begin with the definition of membership inference derived from~\cite{somesh-overfit}.

\begin{experiment}[Membership Inference]
\label{exp:mem-inf}
Let $\adversary$ be the adversary.
\begin{enumerate}
    \item Construct model $f_D$ and give $\adversary$ oracle access to it.
    \item Sample $b \sample{\$} \{0, 1\}$.
    \item Sample $\mathbf{x} \sample{\$} D$ if $b = 0$, else sample $\mathbf{x} \sample{\mathcal{P}} \mathbb{R}^m$. 
    \item $\adversary$ receives $\mathbf{x}$ and continues to make oracle queries to $f_{D}$.
    \item $\adversary$ announces $b' \in \{0, 1\}$. If $b' = b$, output 1, else output 0.
\end{enumerate}
\end{experiment}


Notice that unlike in \cite{somesh-overfit}, we do not necessarily assume that all the members of $D$ are sampled from the same distribution $\mathcal{P}$. Thus, in the above definition, the adversary can possibly gain an advantage in distinguishing a member of the dataset because of its distance from the ``typical population member.'' We also propose a stronger definition below, where the adversary is asked to distinguish between a random member from the dataset and a non-member similar to the member. Such a strong inference attacker, as we show later, has a better chance of inferring missing attributes of a feature vector and violating privacy of individuals in the dataset even when they share ``similar'' features.

\begin{experiment}[Strong Membership Inference]
\label{exp:strong-mem-inf}

Let $\adversary$ be the adversary, and let $r > 0$ be a real number.
\begin{enumerate}
    \item Construct model $f_D$ and give $\adversary$ oracle access to it.
    \item Sample $\mathbf{x}_0 \sample{\$} D$, and $\mathbf{x}_1 \sample{\$} \mathsf{ngb}_r(\mathbf{x}_0) \setminus D$.
    \item $\adversary$ receives $\mathbf{x}_0$ and $\mathbf{x}_1$ (after shuffling), and continues to make oracle queries to $f_{D}$.
    \item $\adversary$ announces $\mathbf{x}'$. If $\mathbf{x}' = \mathbf{x}_0$, output 1, else output 0.
\end{enumerate}
\end{experiment}


The above definition challenges the adversary to distinguish between two neighbouring feature vectors. The closeness of the two vectors is controlled by the parameter $r$ in the definition. We show in Section~\ref{sec:experiment} why such a strong inference attacker is a better starting point for constructing an attribute inference attacker in the spirit of \cite{somesh-overfit}. Next, we present an adapted definition of the attribute inference attacker from \cite{somesh-overfit}.

\begin{experiment}[Attribute Inference]
\label{exp:attr-inf}
Let $\adversary$ denote the adversary, and let $|\phi(\mathbf{x}^*)| = m' \ge 1$.
\begin{enumerate}
    \item Construct model $f_D$ and give $\adversary$ oracle access to it.
    \item Sample $b \sample{\$} \{0, 1\}$.
    \item Sample $\mathbf{x} \sample{\$} D$ if $b = 0$, else sample $\mathbf{x} \sample{\mathcal{P}} \mathbb{R}^m$. 
    \item Let $\mathbf{x}^*$ be the portion of $\mathbf{x}$ such that $|\phi(\mathbf{x}^*)| = m'$.
     \item $\adversary$ receives $\mathbf{x}^*$ and continues to make oracle queries to $f_{D}$.
    \item $\adversary$ announces $\mathbf{x}' \in \mathbb{R}^m$. If $\mathbf{x}' = \mathbf{x}$ output 1, else output 0.
\end{enumerate}
\end{experiment}
The attribute inference advantage of the adversary is the probability that the experiment outputs 1 when $b = 0$ (member of the dataset) minus the probability that the experiment outputs 1 when $b = 1$ (random vector from the population). The above definition mirrors the one from~\cite{somesh-overfit}, except that we do not take the class label into account. 
Our experiments in Section~\ref{sec:experiment} show that constructing an attacker that can \emph{exactly} predict the missing values of a portion of a member vector with high probability is highly unlikely. Thus, we propose below the definition of an approximate AIA, that requires the attacker to predict the missing values only ``approximately close'' to a member vector. 

 \begin{experiment}[Approximate Attribute Inference]
\label{exp:attr-inf-w}

Let $\adversary$ denote the adversary, let $|\phi(\mathbf{x}^*)| = m' \ge 1$, let $\alpha \ge 0$ be a distance parameter.
\begin{enumerate}
   \item Construct model $f_D$ and give $\adversary$ oracle access to it.
    \item Sample $b \sample{\$} \{0, 1\}$.
    \item Sample $\mathbf{x} \sample{\$} D$ if $b = 0$, else sample $\mathbf{x} \sample{\mathcal{P}} \mathbb{R}^m$. 
    \item Let $\mathbf{x}^*$ be the portion of $\mathbf{x}$ such that $|\phi(\mathbf{x}^*)| = m'$.
     \item $\adversary$ receives $\mathbf{x}^*$ and continues to make oracle queries to $f_{D}$.
    \item $\adversary$ announces $\mathbf{x}' \in \mathbb{R}^m$. If $d(\mathbf{x}',\mathbf{x}) \le \alpha$ output 1, else 0.
\end{enumerate}
\end{experiment}
The attribute inference advantage of the adversary is defined analogously to the previous definition.

\section{Experimental Evaluation}
\label{sec:experiment}
\descr{Setup.} We adopt MIA from Salem et al.~\cite{ml-leaks} which directly exploits the maximum confidence score returned by the model for a given input vector. The confidence score is compared against a previously learned threshold and the input is deemed a member or a non-member accordingly. We use two datasets for the evaluation of the attacks: a social network locations check-in dataset obtain from Foursquare (Location)~\cite{yang-location} and a Shopping transactions dataset (Purchase).\footnote{\url{https://www.kaggle.com/c/acquire-valued-shoppers-challenge/data}} Both datasets have been used previously to demonstrate MIA~\cite{shokri-mia, ml-leaks}. The datasets are binary, with 467 binary features in Location and 699 in Purchase. The class labels in both the Location and Purchase datasets are obtained through k-means clustering. There are 30 classes in the Location dataset, and 5 variants of the Purchase dataset differing in the number of classes (2, 10, 20, 50, 100) as is done in~\cite{ml-leaks}. For attack success, we report the Area Under the ROC (Receiver Operating Characteristic) Curve (AUC) which is obtained by varying the threshold between extremes. 
The AUCs are obtained by sampling up to 10,000 vectors from the dataset (Our location dataset only contains 6,951 vectors in total, whilst the purchase dataset contains a total of 200,000 vectors), then splitting the dataset into 20/80 for training/testing respectively.
From the training and testing sets, we sample 1000 vectors each to create member and non-member sets, respectively. These subsamples are evaluated by the target model for maximum confidence values, which are used to compute the AUC.
The target model is trained locally with Tensorflow Estimators\footnote{\url{https://www.tensorflow.org/guide/estimators}} as a fully connected neural network with 5 hidden layers of [1024, 1024, 512, 512, 512] nodes for all Purchase datasets, and [512, 512, 512, 512, 512] nodes for Location. This configuration reproduces MIA AUCs reported in~\cite{ml-leaks}: we obtained AUCs averaged over 50 iterations as 0.872 for Location-30, and 0.548, 0.628, 0.671, 0.745, 0.794 for Purchase-2, 10, 20, 50, 100 respectively. These illustrate membership inference as per Definition \ref{exp:mem-inf}.

\medskip
In the following, we first show that the Hamming distance of non-members from every vector of the training dataset in the above experiments is considerably large, owing to the success of MIA. We then construct synthetic non-members by changing the (binary) feature values to control the Hamming distance from the training dataset, and show that if the Hamming distance is low, the AUCs are close to random guess (0.5) and only improve as the Hamming distance increases. We follow this up with implications to attribute inference (Definition~\ref{exp:attr-inf}), and evaluate the performance of the approximate attribute inference attack (Definition~\ref{exp:attr-inf-w}). 

\begin{figure}[t!]
	\centering
    \includegraphics[width=1.0\columnwidth]{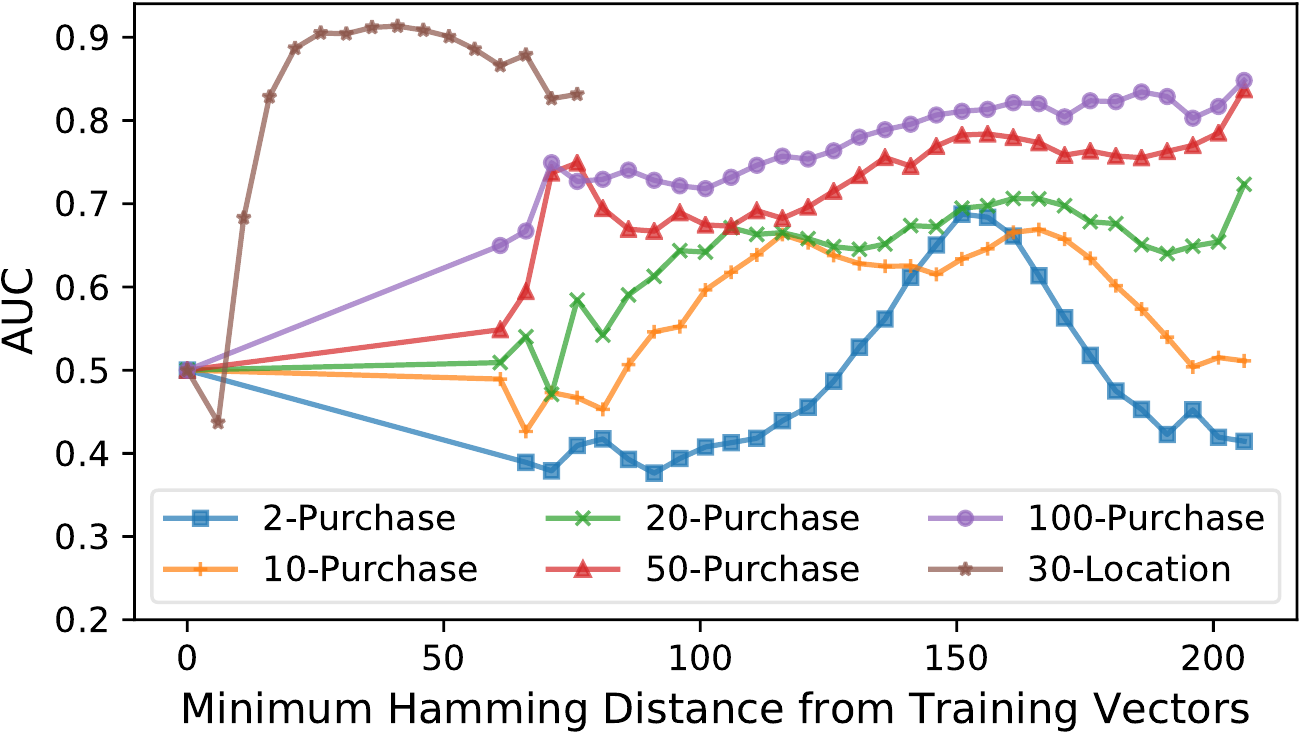}
    \vspace{-7mm}
  	\caption{Increasing AUC of MIA with increasing Hamming distance of actual non-members from the training dataset.}
 	\label{fig:nmvec_auc}
\end{figure}

\descr{MIA Performance on the Testing Set.} After training the target model, we compute the Hamming distance of each non-member vector from the training set (recall from Section~\ref{sec:definition} that this is the minimum Hamming distance from any member). The vectors are then grouped according to the Hamming distances from the training dataset (note that the Hamming distance is 0 for members). We then calculate AUC for each Hamming distance grouping. This test is repeated 50 times, and the AUC is computed on the aggregation of all confidence values (Figure~\ref{fig:nmvec_auc}). For the Location dataset, as hypothesized, the AUC is close to random guess (0.5) for non-members close to the training dataset, and starts improving as we shift away. This trend is also visible for the Purchase datasets, although less so. This is because for the Purchase dataset the non-members are farther away from the training dataset. This shows that strong membership inference (Definition~\ref{exp:strong-mem-inf}) is less successful than the reported high accuracy of MI that falls under Definition~\ref{exp:mem-inf}. 

\begin{figure}[t!]
	\centering
    \includegraphics[width=1.0\columnwidth]{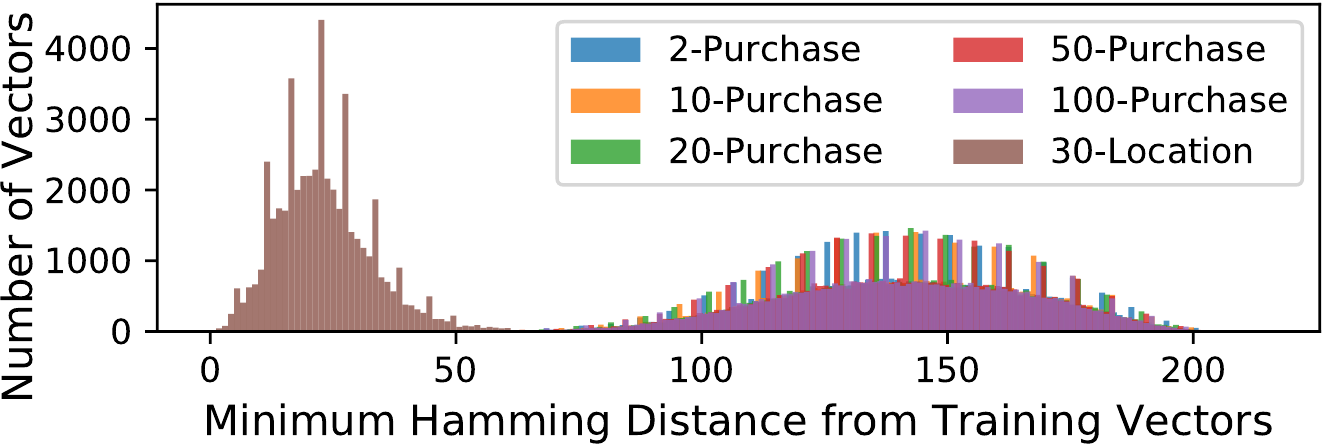}
    \vspace{-7mm}
  	\caption{Histogram of Hamming distances of non-members from different training datasets.}
 	\label{fig:nmvec_hist}
 	\vspace{-1mm}
\end{figure}

The lack of vectors close to and farthest away from the training dataset is due to the distribution of Hamming distances displayed in Figure \ref{fig:nmvec_hist}. As the non-members within the original Purchases dataset do not provide a full picture of how the AUC, and hence MIA performance, changes at small Hamming distances from members, we generate artificial vectors.

\begin{figure}[t!]
	\centering
    \includegraphics[width=1.0\columnwidth]{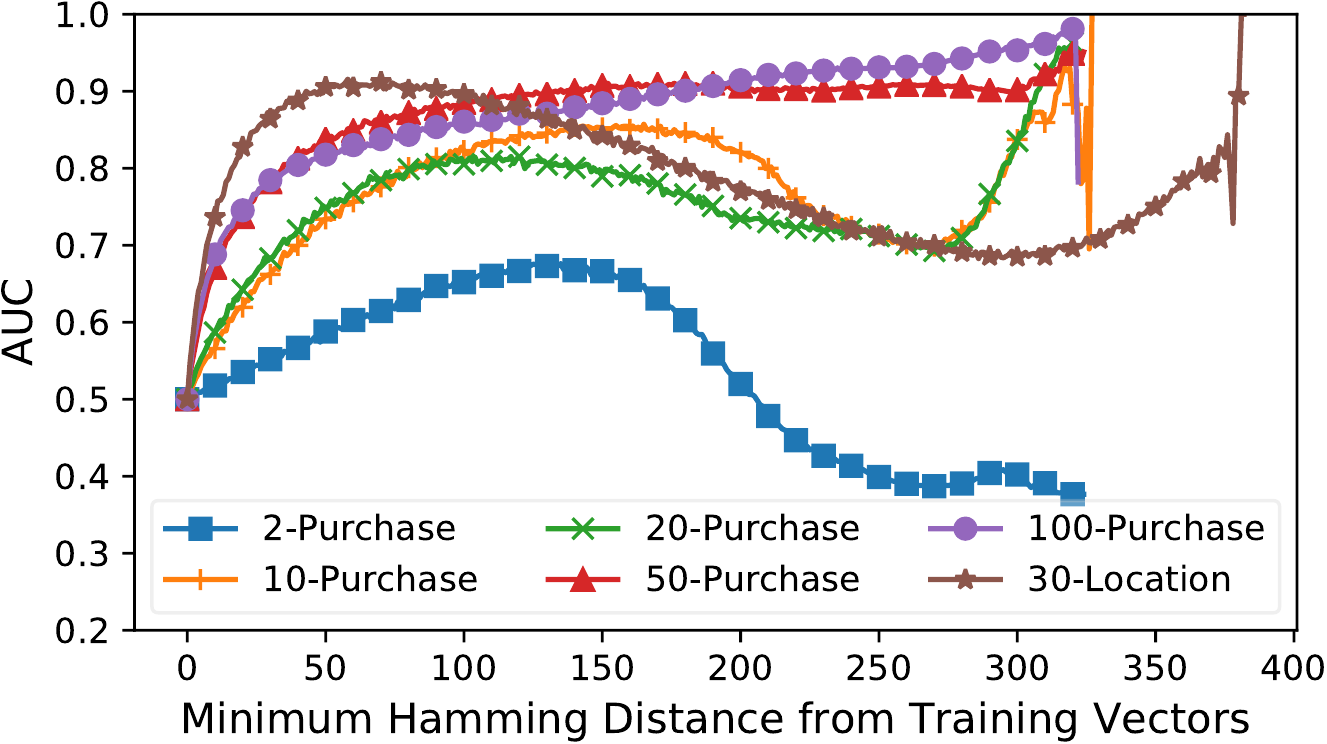}
    \vspace{-7mm}
  	\caption{Increasing AUC of MIA with increasing Hamming distance of synthetic non-members from the training dataset.}
 	\label{fig:genvec_auc}
 	\vspace{-3mm}
\end{figure}

\descr{MIA Performance on Synthetic Vectors Close to Members.}
To generate new vectors, we (a) randomly select a member of the training set, and (b) randomly select features to invert. We vary features until a minimum number of variants (5) are produced for each Hamming distance group. This is repeated for all selected 1000 member vectors. The vectors thus generated are non-members (unless, by chance, any of them collides with a member, in which case we discard it). We then compute the AUC of MIA displayed in Figure \ref{fig:genvec_auc}. As can be seen, the AUC is close to 0.5 for non-members close to the training dataset, and starts improving as the distance from training dataset increases. Also, the higher the number of classes, the steeper the improvement in AUC as the Hamming distance increases. Interestingly, in the Purchase datasets, for smaller number of classes (2, 10 and 20), we observe an increase an AUC, followed by a decrease. For the 10 and 20 class variant, we see a second incline around a Hamming distance of 250. We see the same trend in the Location dataset but at different Hamming distance \blue{of 300}. This may be related to the distribution of non-member vectors in the original dataset. We plan to investigate this in the future.

\begin{figure}[t!]
	\centering
    \includegraphics[width=1.0\columnwidth]{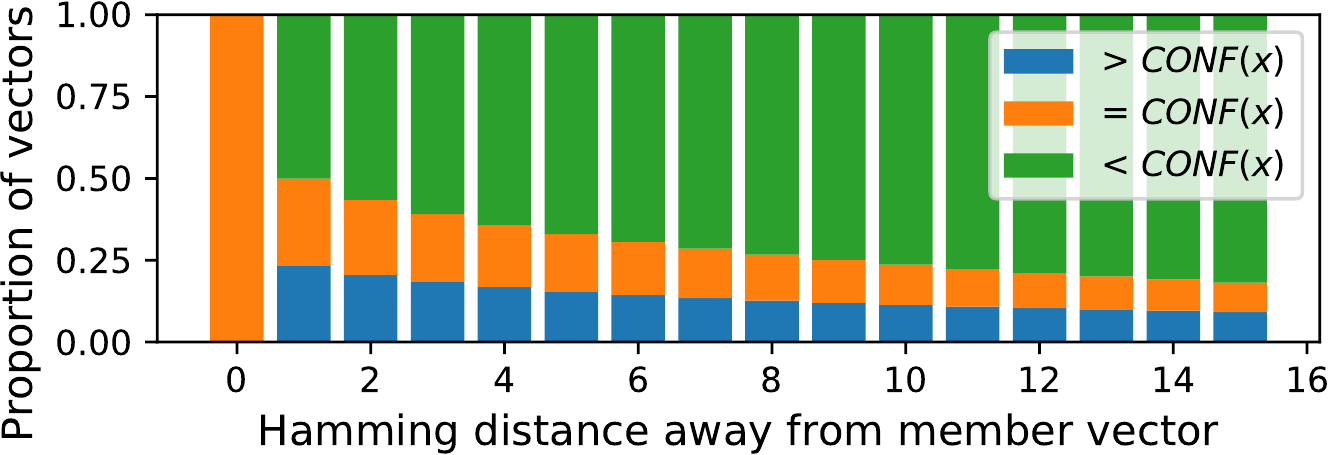}
    \vspace{-7mm}
  	\caption{Permuted vectors from an original vector $\mathbf{x}$ with larger hamming distances are more likely to have lower confidence values than $\mathbf{x}$. $\mathbf{CONF(x)}$ is the confidence value of $\mathbf{x}$.}
  	\label{fig:mrmr_attrinf}
\end{figure}

\descr{Attribute Inference.}
The above results indicate that if an adversary has a portion $\mathbf{x}^*$ of a feature vector $\mathbf{x}$ with a single unknown feature having value $b$ in the training dataset, then it is difficult to distinguish between the (member) vector $\mathbf{x} = (\mathbf{x}^*, b)$ and $(\mathbf{x}^*, \overline{b})$. Similarly, if the unknown part of $\mathbf{x}^*$ is bigger, say 15 unknown features, and only one vector $\mathbf{x}$ exists in the training dataset with the portion $\mathbf{x}^*$, then it is difficult to distinguish between $\mathbf{x}$ and vectors \emph{close} to $\mathbf{x}$ having the same portion. We demonstrate this with the location dataset in Figure \ref{fig:mrmr_attrinf}, which shows the distribution of confidence values with respect to Hamming distances from member vectors. Confidence values of non-member vectors that are a small Hamming distance away from member vectors are almost equally likely to be higher or lower than those of member vectors. It is due to this reason that we propose a relaxed version of attribute inference (Definition~\ref{exp:attr-inf-w}) where we allow the adversary to guess a vector close to the target vector. Next we analyze the performance of this approximate AI attack.

\begin{figure}[t!]
	\centering
    \includegraphics[width=1.0\columnwidth]{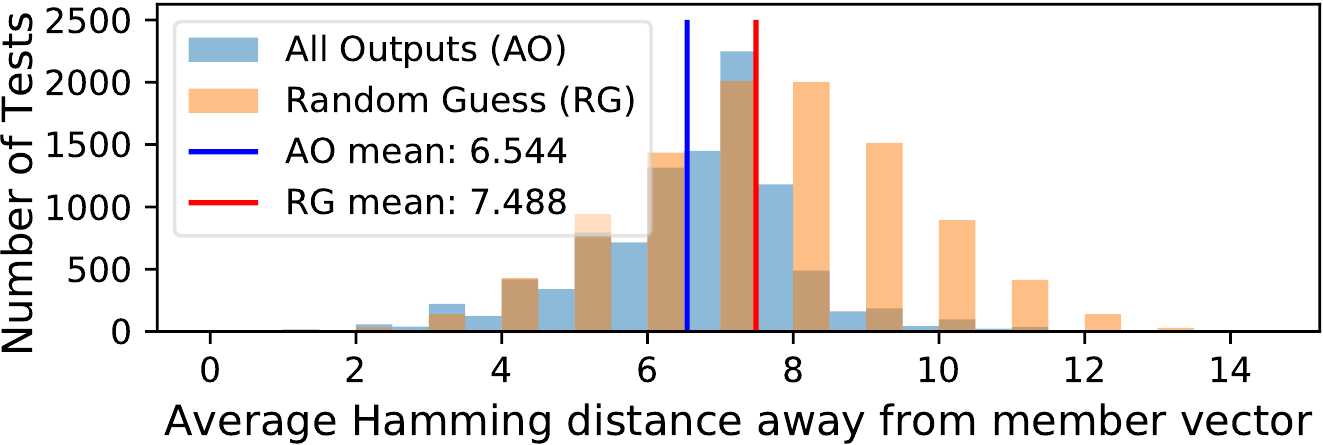}
    \vspace{-7mm}
  	\caption{Distribution of average Hamming distances of non-member vectors from member vectors with the same portion.}
 	\label{fig:mrmr_attrinf1}
 	\vspace{-3mm}
\end{figure}

To give attribute inference the highest likelihood of success, we use the 15 most important features as the unknown part of $\mathbf{x}^*$, ranked according to the minimal-redundancy-maximal-relevance (mRMR) criteria~\cite{peng-mrmr}. A feature with the highest mRMR ranking should statistically contribute the most information about the class label, and hence should observe the largest deviations in prediction confidences. \blue{For each target member in the training dataset,} 
\blue{the approximate AI attack algorithm generates} all possible permutations of the 15 unknown features (only one of which corresponds to the original vector $\mathbf{x}$). The algorithm then returns the vector with the highest confidence value as its guess for the target vector (and hence the missing attributes). We test this on the Location dataset and plot the results in Figure ~\ref{fig:mrmr_attrinf1}. The plot shows the distribution of Hamming distances of the outputs of the attack to the member vectors. In the case of a tie (multiple vectors yielding the same maximum confidence score), we report the average Hamming distance of the tied vectors to the member vector. Randomly guessing the unknown part would result in a mean Hamming distance of 7.5. The plot shows that the attribute inference attack performs considerably better with a mean distance of approximately 6.5, with the bulk of the distribution being closer to member vector.





\section{Related Work}
Many privacy researchers have identified the issue of membership inference in machine learning models. Shokri et al.~\cite{shokri-mia} propose the idea of constructing \emph{shadow models} which mimic the unknown training datasets to launch an MIA. This technique is based on a related observation to ours that machine learning models trained on \emph{similar} data behave similarly. One method to generate data similar to the training dataset in~\cite{shokri-mia} is to flip 10 to 20\% of the (binary) features in the dataset. Salem et al.~\cite{ml-leaks} simplify MIA by dropping the need for constructing (the rather expensive) shadow models. In particular, they demonstrate that the difference in confidence scores between members and non-members is enough to distinguish them. Indeed, we have adopted this attack to demonstrate the relationship between membership and attribute inference, and the distance from member vectors. Both works have used a split of a real dataset into training and testing sets, and demonstrated the effectiveness of MIA using the testing sets. We have shown that most vectors in the testing set, i.e., non-members, are expected to be far from the training set, which explains why the relationship of MIA to distance from members was not identified in these works.  

Somesh et al.~\cite{somesh-overfit} formally relate overfitting to the effectiveness of MIA, a link which was previously experimentally identified and demonstrated in~\cite{shokri-mia, ml-leaks}. They also formally define attribute inference (which is the basis for our related definition). However, their definition as well as experimental evaluation of attribute inference uses the class label as well. An interesting area of future research is to see how inclusion of class label in our strong membership and approximate attribute inference definitions influence the result. A related attack on machine learning models is \emph{model extraction}~\cite{tramer-stealing}, through which unknown parameters of the model are retrieved to construct similarly behaving models (hence stealing the model in a proprietary sense). These attacks are applicable to the entire model itself and not necessarily related to individuals in the training dataset. Related to above is the question whether MIA can identify \emph{biases} in the training datasets. This has been demonstrated in~\cite{disparate-sub-group}, in which the authors show that even if MIA is ineffective as a whole on a dataset, it has disparate effectiveness on different sub-groups in the dataset. This in turn reveals some information about the distribution of the underlying dataset. In some sense the shadow model technique also relies on the fact that machine learning models may not be able to suppress some characteristics of the dataset (such as its overall distribution).
%
We have defined Definition~\ref{exp:mem-inf} and~\ref{exp:attr-inf} such that the attacker may be able to distinguish \blue{between a member and a non-member} via the bias in the training dataset distribution; which, depending on the application, may be construed as a privacy violation. 

\section{Future Work}
We plan to further refine the inference definitions, investigate algorithmic approaches to infer attributes in the weak attribute inference model, and analyze the effect of over-fitting on the results. We would also like to see whether these results are replicated by other classifiers, and by other (non-binary) datasets. With non-binary datasets, our definition will be expanded to explore other distance metrics such as Euclidean distance. We will also investigate if attribute inference can be performed as a standalone attack, i.e., without using membership inference as a subroutine. 

\bibliographystyle{ACM-Reference-Format}
\bibliography{references}

\end{document}